# PHASE–TRANSITION THEORY OF INSTABILITIES.
## II. FOURTH–HARMONIC BIFURCATIONS AND λ–TRANSITIONS

Dimitris M. Christodoulou[1], Demosthenes Kazanas[2], Isaac Shlosman[3,4], and Joel E. Tohline[5]

## ABSTRACT

We use a free–energy minimization approach to describe in simple and clear physical terms the secular and dynamical instabilities as well as the bifurcations along equilibrium sequences of rotating, self–gravitating fluid systems. Our approach is fully nonlinear and stems from the Landau–Ginzburg theory of phase transitions. In this paper, we examine fourth–harmonic axisymmetric disturbances in Maclaurin spheroids and fourth–harmonic nonaxisymmetric disturbances in Jacobi ellipsoids. These two cases are very similar in the framework of phase transitions.

It has been conjectured (Hachisu & Eriguchi 1983) that third–order phase transitions, manifested as smooth bifurcations in the angular momentum–rotation frequency plane, may occur on the Maclaurin sequence at the bifurcation point of the axisymmetric one–ring sequence and on the Jacobi sequence at the bifurcation point of the dumbbell–binary sequence. We show that these transitions are forbidden when viscosity maintains uniform rotation. The uniformly rotating one–ring/dumbbell equilibria close to each bifurcation point and their neighboring uniformly rotating nonequilibrium states have higher free energies than the Maclaurin/Jacobi equilibria of the same mass and angular momentum. These high–energy states act as free–energy barriers preventing the transition of spheroids/ellipsoids from their local minima to the free–energy minima that exist on the low rotation frequency branch of the one–ring/binary sequence. At a critical point, the two minima of the free–energy function are equal signaling the appearance of a *first–order* phase transition. This transition can take place beyond the critical point only nonlinearly if the applied perturbations contribute

[1] Virginia Institute for Theoretical Astronomy, Department of Astronomy, University of Virginia, P.O. Box 3818, Charlottesville, VA 22903

[2] NASA Goddard Space Flight Center, Code 665, Greenbelt, MD 20771

[3] Gauss Foundation Fellow

[4] Department of Physics & Astronomy, University of Kentucky, Lexington, KY 40506

[5] Department of Physics & Astronomy, Louisiana State University, Baton Rouge, LA 70803

enough energy to send the system over the top of the barrier (and if, in addition, viscosity maintains uniform rotation).

In the angular momentum–rotation frequency plane, the one–ring and dumbbell–binary sequences have the shape of an "inverted S" and two corresponding turning points each. Because of this shape, the free–energy barrier disappears suddenly past the higher turning point leaving the spheroid/ellipsoid on a saddle point but also causing a "catastrophe" by permitting a "secular" transition toward a one–ring/binary minimum energy state. This transition *appears* as a typical second–order phase transition although there is no associated sequence bifurcating at the transition point (cf. Paper I).

Irrespective of whether a nonlinear first–order phase transition occurs between the critical point and the higher turning point or an apparent second–order phase transition occurs beyond the higher turning point, the result is fission (i.e. "spontaneous breaking" of the *topology*) of the original object on a secular time scale: the Maclaurin spheroid becomes a uniformly rotating axisymmetric torus and the Jacobi ellipsoid becomes a binary. The presence of viscosity is crucial since angular momentum needs to be redistributed for uniform rotation to be maintained.

We strongly suspect that the "secular catastrophe" is the dynamical analog of the notorious $\lambda$–transition of liquid $^4$He because it appears as a "second–order" phase transition with infinite "specific heat" at the point where the free–energy barrier disappears suddenly. This transition is not an elementary catastrophe. In contrast to this case, a "dynamical catastrophe" takes place from the bifurcation point to the lower branch of the Maclaurin toroid sequence because all conservation laws are automatically satisfied between the two equilibrium states. Furthermore, the free–energy barrier disappears gradually and this transition is part of the elementary *cusp catastrophe*. This type of "$\lambda$–transition" is the dynamical analog of the Bose–Einstein condensation of an ideal Bose gas.

The phase transitions of the dynamical systems are briefly discussed in relation to previous numerical simulations of the formation and evolution of protostellar systems. Some technical discussions concerning related results obtained from linear stability analyses, the breaking of topology, and the nonlinear theories of structural stability and catastrophic morphogenesis are included in an Appendix.

*Subject headings:*   galaxies: evolution – galaxies: structure – hydrodynamics – instabilities – stars: formation

## 1   INTRODUCTION

### 1.1   General Remarks

We continue exploring the dynamics of bifurcations and instabilities along equilibrium sequences of rotating, self–gravitating, incompressible, fluid systems using the energy–minimization approach that we introduced in the first paper of this series (hereafter referred to as Paper I). This fully nonlinear approach is based on the theory of phase transitions of Landau & Ginzburg (see e.g. Landau & Lifshitz 1986) and aims toward understanding the physics of points of bifurcation and instability. Such points may be better understood as phase–transition points beyond which a fluid (or stellar) system evolves away from its original equilibrium state and begins an approach toward a new equilibrium state of lower free energy. The theory of phase transitions was originally introduced by Bertin & Radicati (1976) for the Maclaurin–Jacobi bifurcation point and was later extended to more points by group–theoretical methods (Constantinescu, Michel, & Radicati 1979) and with the help of numerical computations (Hachisu & Eriguchi 1983, 1984a). It is also clear from the general descriptions of secular and dynamical stability of Lamb (1932), Lyttleton (1953), and Lebovitz (1977) that the power and simplicity of this approach had been appreciated, but not utilized or properly exploited, in the past.

In Paper I, we saw that the second–harmonic secular and dynamical instabilities on the Maclaurin sequence of oblate spheroids can be understood as second–order phase transitions taking place toward the Jacobi or Dedekind sequence and toward the $x=+1$ self–adjoint Riemann sequence, respectively. The distinguishing characteristic that decides which of the three phase transitions is allowed to take place is the (non)–conservation of a particular integral of motion: if both angular momentum and circulation are conserved during evolution, then only the dynamical instability toward the $x = +1$ self–adjoint Riemann sequence is relevant; if only angular momentum (or only circulation) is conserved, then only the secular instability toward the Jacobi (Dedekind) sequence is relevant.

The phase–transition time scales are also different between cases. The second–order phase transition toward the $x=+1$ sequence is dynamical because angular momentum and circulation are both automatically conserved between the initial and the final equilibrium state. More importantly, there also exist intermediate nonequilibrium states that a system passes through in which both integrals of motion are also conserved. [In the past, this crucial point has not been given the attention it deserves. We shall see in the third paper of this series (hereafter referred to as Paper III) that this simple observation is elevated to a role of fundamental importance in discussing evolution on the Jacobi sequence and the fission problem in general.] The second–order phase transition toward the "fluid" Jacobi (or Dedekind) sequence is secular because circulation (angular momentum) conservation is destroyed slowly by the action of viscosity (gravitational radiation or any other angular momentum loss mechanism that conserves circulation). In stellar systems, the second–order phase transition toward the "stellar" Jacobi sequence is dynamical because the circulation is destroyed on a dynamical time scale by the off–diagonal components of the stress–tensor gradient (terms in the equations of motion having the same order of magnitude as the conventional "pressure" gradients).

In this paper, we examine the energetics of two more bifurcations in fluids that bring into existence the one–ring sequence and the dumbbell–binary sequence. The one–ring sequence bifurcates from the Maclaurin sequence at a point where the meridional eccentricity of the spheroid is $e=0.985226$ (Chandrasekhar 1967a; see also Bardeen 1971). The properties of equilibria along this sequence have been computed by Eriguchi & Sugimoto (1981) and by Hachisu, Eriguchi, & Sugimoto (1982). The one–ring sequence is initially composed of equilibria that look like "concave hamburgers" and continues on with purely toroidal equilibria. The dumbbell–binary sequence bifurcates from the Jacobi sequence at a point where $e$=0.9663 and the equatorial eccentricity is $\eta$ = 0.9548. These values correspond to ratios $b/a$=0.2972 and $c/a$=0.2575 of the axes $a, b, c$ of the Jacobi ellipsoid (Chandrasekhar 1969, hereafter referred to as EFE; see also Chandrasekhar 1967b, 1971). The properties of equilibria along this sequence have been computed by Eriguchi, Hachisu, & Sugimoto (1982). This sequence is initially composed of equilibria that look like "dumbbells" and continues on with detached equal–mass binaries.

We analyze these two sequences together because their dynamical properties are very similar (almost identical) in the framework of phase transitions (Hachisu & Eriguchi 1983). Understanding the intricacies related to the one–ring sequence is important for the following reason. The phase transitions that appear between the Maclaurin and the one–ring sequence and other related transitions have been studied in the past but they have not been understood entirely (Hachisu & Eriguchi 1983; Tohline 1985; Tohline & Christodoulou 1988; Christodoulou & Tohline 1990). None of these studies has pointed out the following discrepancy: Bardeen (1971) found that dynamical axisymmetric instability sets in on the Maclaurin sequence at $e$=0.99892. The point of dynamical instability does not appear to be related to the intricate structure of the one–ring sequence that has the shape of an "inverted S" and thus two turning points in the angular momentum–rotation frequency plane (see Figure 1 in §2 below). In fact, the point $e$=0.99892 lies way beyond the bifurcation point $e$=0.99375 of the two–ring sequence which was found by Eriguchi & Hachisu (1982). Furthermore, Hachisu, Tohline, & Eriguchi (1987) argued that the point $e$=0.99892 is not even the bifurcation point of the so–called Maclaurin toroid sequence (another ring–like sequence discussed in detail by Eriguchi & Hachisu 1985; see also §4.2 and Figures 6, 7 below), as was originally believed. If that were true, Bardeen's point of dynamical axisymmetric instability would be unrelated to any known equilibrium sequence.

On the other hand, the phase–transition studies cited above did point out (in different ways) that a "dynamical" instability, manifested as some kind of phase transition, does appear between the Maclaurin sequence and a sequence of toroidal equilibria. As we shall see below, the "inverted S" shape of the one–ring sequence is responsible for a first–order as well as an apparent second–order phase transition. However, these transitions may only take place between uniformly rotating states. Thus, viscosity is implicitly assumed to maintain uniform rotation by redistributing angular momentum efficiently. Because the transitions are effectively driven by the action of viscosity they correspond to *secular instabilities* — although they may be quite fast since they can occur only when strong viscosity is present. The apparent second–order phase transition is directly related to the higher turning point of the one–ring sequence. For this point, we have obtained $e = 0.986834$ on the Maclaurin sequence based on the angular momentum value given by Eriguchi & Sugimoto (1981). Understanding where Bardeen's (1971) dynamical instability fits in the above picture is not a trivial task. In §4.2, we explain that this instability must appear at the bifurcation point of the Maclaurin toroid sequence.

Understanding the phase transitions from the Maclaurin to the one–ring sequence as well as similar transitions from the Jacobi to the dumbbell–binary sequence is also impor-tant because the dynamics involved is extremely rich although it takes place in the comparatively simple setting of uniformly rotating, self–gravitating, incompressible systems. We have thus an opportunity to investigate, visualize, and hopefully better understand physical processes that are quite common but also quite complicated in other branches of physics, such as fundamental interactions, spontaneous symmetry breaking, the $\lambda$–transition of liquid $^4$He, and structural instabilities/catastrophes (see related discussions in the Appendix). For our incompressible–fluid models, we show in §2 and in §3 that the above pairs of "mother–daughter" sequences do not "communicate" naturally through the bifurcation points. Specifically, third–order phase transitions precisely at the bifurcation points (Hachisu & Eriguchi 1983) are strictly forbidden on energetic grounds. A phase transition does appear in the vicinity of each bifurcation point but it is of first order and free–energy barriers separate the two minima of the free–energy function. Under these circumstances, communication can be achieved only if nonlinear perturbations provide enough extra energy to drive a system away from the "mother" sequence, over the top of the barrier, and finally down toward the new equilibrium state that belongs to the "daughter" sequence.

On the other hand, the barrier that separates the two stable equilibrium states disappears suddenly past the higher turning point of each daughter sequence in which case a phase transition is allowed to take place in the presence of viscosity. It is this phase transition which appears to be of second order although there is no bifurcating sequence analogous to the Jacobi or to the Dedekind sequence. As a result, the "second–order" phase transition takes place between sequences "discontinuously" in the sense that a quantity analogous to the specific heat $c_v$ diverges at the transition point (see also Paper I). This is understood because the transition takes place between states of different rotation frequency $\Omega$ but at constant angular momentum $L$ (see also §1.2 below for a discussion of all the assumptions involved). As a result, $c_v \propto d\Omega/dL \to \infty$ at the transition point (Paper I) and returns to a finite value as soon as the evolving system reaches the daughter sequence. This behavior of the analog to the specific heat is certainly also related to the fact that no "symmetry breaking" results from this transition — only the topology "breaks" when a spheroid becomes a torus or when an ellipsoid becomes a binary. (We have discussed "topology breaking" in Paper I and we shall return to it in the Appendix.)

As was mentioned above, we have studied similar types of phase transitions in the past (e.g. Tohline 1985; Tohline & Christodoulou 1988; Christodoulou, Sasselov, & Tohline 1993) but we were not familiar with the work of Hachisu & Eriguchi (1983) at that time. We did realize, however, that there are two distinct phases of evolution (just as we described above) and we used to call them the " first–order phase transition" and the "Jeans instability" in the context of protostellar collapse (see e.g. Tohline 1985; Christodoulou & Tohline 1990). On the other hand, the results presented in §2 and §3 below strongly suggest that the latter phase is the dynamical analog of the $\lambda$–transition of liquid $^4$He (e.g. London 1954; Huang 1963; Wilks & Betts 1987) which has similar macroscopic properties (i.e. $c_v \to \infty$ at the $\lambda$–point and apparently no symmetry breaking). For this reason, we shall heretofore call this type of phase transition between a mother and its daughter sequence the "$\lambda$–transition" by analogy to the superfluid transition. We note that the "secular" $\lambda$–transitions of §2 and §3 are effectively driven by strong viscosity, so they are different than the "dynamical" $\lambda$–transition of §4.2 for which the name "Jeans instability" is appropriate.

## 1.2   Basic Assumptions and Dimensionless Quantities

In what follows, we consider for simplicity only the evolution of incompressible, uniformly rotating fluid–masses that contract under the action of self–gravity. We also adopt the following assumptions: (1) Mass is always strictly conserved not only between the initial state and the final equilibrium state but also in the intermediate states that an evolving fluid–mass passes through. (2) Energy always decreases as a fluid–mass evolves toward a new equilibrium state. (3) Angular momentum, expressed in physical units, is always strictly conserved but, in normalized units, it increases or stays constant in time. This general increase in normalized angular momentum $j$ reflects a general increase of the density $\rho$ during the evolution of a contracting fluid–mass [$j \propto \rho^{1/6}$ in equation (1.2) below]. (4) Circulation is strictly conserved only in the complete absence of "viscosity" (i.e. in a "perfect" fluid) but its absolute value decreases during the evolution of a "viscous" fluid–mass.

The transitions between equilibrium sequences discussed in §2 and §3 are "discontinuous" and "secular" in the sense that they do not occur at traditional bifurcation points. In each case, a daughter sequence with a branch of lower free energy has already bifurcated from the mother sequence at a previous point. Thus, a lower energy state already exists but a transition is prevented by a free–energy barrier that separates the two stable equilibria. As soon as this barrier disappears, a "secular" transition is allowed to take place from a saddle point on the mother sequence to a lower energy state on the daughter sequence. Since both the initial and the final equilibrium state are assumed to be in uniform rotation, the transition is effectively driven by viscosity which must then be assumed sufficiently strong to enforce uniform rotation during evolution.

Furthermore, the transitions discussed in this paper preserve symmetry. [The general case of both symmetry and topology breaking is relevant to the results presented in the fourth paper of this series (hereafter referred to as Paper IV) and is briefly discussed in §4.3 below.] They are then allowed to proceed at constant mass $M$ and constant angular momentum $L$. Although the free–energy function $E$ cannot be derived in closed form, we can still describe its qualitative features based on the conservation laws.

Consider for simplicity only $\lambda$–transitions between axisymmetric sequences. In this case, $E$ is a function of four variables, the rotation frequency $\Omega$ and three length scales $a$, $c$, and $R_o$; because symmetry is preserved, $E$ does not depend on a vorticity variable $x$ as in Paper I. The length scales $a$ and $c$ represent the equatorial and polar axes just as in Maclaurin spheroids. The third length scale $R_o$ must be introduced here to account for topology breaking (cf. Paper I) — a common characteristic of all $\lambda$–transitions (see Appendix). For example, in a torus, $R_o$ represents the size of the (non)–equilibrium system while, in a "concave hamburger," it represents a third point on the distorted surface of the (non)–equilibrium configuration where the pressure is also zero (i.e. in addition to the end–points of the axes $a$ and $c$). Assuming that the mass $M$ and the angular momentum $L$ are conserved along an entire evolutionary path, we can then eliminate from $E$ only two variables in favor of $M$ and $L$. $E$ becomes then a function of two independent variables and depends on two integrals of motion.

Our interpretations of points of instability and bifurcation rely on the numerical work of Eriguchi, Hachisu, and Sugimoto who published their results in a series of fifteen papers between 1981 and 1986. These works are cited at the end of this paper. We refer below to particular results from these numerical computations as we need them. We also adopt for convenience the dimensionless quantities $\omega$ (rotation frequency), $j$ (angular momentum), and $E$ (total energy) that were used by the authors in the presentation of their numerical results. These quantities are defined by

$$\omega^2 \equiv 10^2 \Big( \frac{\Omega^2}{4\pi G \rho} \Big), \tag{1.1}$$

$$j^2 \equiv 10^2 \Big( \frac{L^2}{4\pi G \rho^{-1/3} M^{10/3}} \Big), \tag{1.2}$$

and

$$E \equiv 10^4 \Big[ \frac{T+W}{(4\pi G)^2 M^5 L^{-2}} \Big], \tag{1.3}$$

where $\Omega$ is the rotation frequency, $G$ is the gravitational constant, $\rho$ is the mass density, $L$ is the total angular momentum, and $M$ is the mass, all in physical units. The term $T+W$ denotes the total energy in physical units as the sum of the total kinetic energy due to rotation $T$ and the total gravitational potential energy $W$. As we have seen in Paper I, the total energy $T+W$ and its dimensionless counterpart $E$ may also represent the free-energy function when they are not constrained by equilibrium conditions (for more details see Tohline & Christodoulou 1988). Finally, when $\omega$, $j$, and $E$ refer specifically to equilibrium configurations, we write them as $\omega_o$, $j_o$, and $E_o$, respectively.

### 1.3  Outline

The remainder of the paper is organized as follows. In §2, we analyze the dynamics and the energetics of the phase transitions that appear between the Maclaurin and the one–ring sequence. Similarly, in §3, we briefly account for the fact that the same types of phase transitions appear between the Jacobi and the dumbbell–binary sequence. In §4, we summarize our results and we relate them to previous numerical simulations of evolving protostellar models. A discussion of additional technical topics — including the linear stability analyses of Chandrasekhar (1967a, b, 1971) and Bardeen (1971), and applications of catastrophe theory and to superfluids — are deferred to an Appendix so that the main text contains only the results which are relevant to star formation studies.

## 2  From the Maclaurin to the One–Ring Sequence

### 2.1  Previous Work

Bifurcation points along the Maclaurin sequence corresponding to various harmonics have been determined by applying tensor virial techniques (Chandrasekhar 1967a, EFE), series expansions of spheroidal wavefunctions for the gravitational potential (Bardeen 1971; Hachisu & Eriguchi 1984c; Eriguchi & Hachisu 1985), and group–theoretical techniques (Constantinescu, Michel, & Radicati 1979).

Hachisu & Eriguchi (1984c) have summarized their analytical and numerical work on the incompressible Maclaurin sequence in one plot that shows most of the following bifurcating sequences: the Jacobi sequence was studied by Hachisu & Eriguchi (1982); the triangle, square, and ammonite sequences were studied by Eriguchi & Hachisu (1982); the one–ring sequence was studied by Eriguchi & Sugimoto (1981), by Hachisu, Eriguchi, & Sugimoto (1982), and by Hachisu (1986a); and the two–ring and core–ring sequences were studied by Eriguchi & Hachisu (1983a). In addition, another sequence of toroidal equilibria, called the Maclaurin toroids, was studied extensively by Eriguchi & Hachisu (1985) and by Hachisu, Tohline, & Eriguchi (1987).

FIGURE 1. *The rotation frequency squared $\omega_o^2$ along the one–ring and the Maclaurin sequence is plotted as a function of the dimensionless angular momentun squared $j_o^2$. The Maclaurin sequence has been calculated analytically. The bifurcation point (filled circle) has also been calculated analytically using a meridional eccentricity of $e$=0.985226 (Chandrasekhar 1967a). All points (open circles) of the one–ring sequence other than the bifurcation point were obtained from Table I of Eriguchi & Sugimoto (1981). Letter symbols indicate points discussed in the text. The dotted line XBC denotes the onset of secular instability.*

Hachisu & Eriguchi (1983) were the first to point out that the Maclaurin ("mother") sequence and the one–ring ("daughter") sequence may be communicating in two different ways which can be thought as phase transitions of third and first order. However, they did not examine the behavior of the free–energy function out of equilibrium near the transition points. Indeed, such an examination reveals the existence of two allowed phase transitions, but these transitions are substantially more complicated and proceed in different ways than previously conjectured. By monitoring the behavior of the free–energy function, we show below that a third–order phase transition at the bifurcation point is strictly forbidden and that one of the two allowed transitions is quite unusual: it does not fit the standard classification scheme of thermodynamical phase transitions, i.e. it cannot be identified with a particular order related to a discontinuity in a particular derivative of the free–energy function.

### 2.2   The One–Ring Sequence

For the purposes of our analysis, we plot in Figures 1 and 2 the one–ring sequence in the region where it bifurcates from the Maclaurin sequence in the $(j_o^2, \omega_o^2)$ and the $(j_o^2, E_o)$ plane, respectively. These plots are similar to Figures 1 and 2 of Hachisu & Eriguchi (1983), but the data plotted here come from Eriguchi & Sugimoto (1981). In any case, this difference does not modify our interpretations below. All of our conclusions can be obtained by also analyzing Figures 1 and 2 of Hachisu & Eriguchi (1983) instead of our figures.

FIGURE 2. *The free energy $E_o$ is plotted versus $j_o^2$ for the sequences of Figure 1.*

Notice in Figure 1 that the one–ring sequence has the shape of an "inverted S" and thus two turning points denoted as points B and D. We call B the higher and D the lower turning point. Eriguchi & Hachisu (1983a) have discussed the turning points that appear in various axisymmetric multi–ring sequences. Tohline (1985) has also discussed such turning points and the related phase transitions in connection with the collapse of axisymmetric, protostellar clouds. An analogous "straight S" shape occurs in the theory of dwarf novae outbursts (the variable viscosity model) where it is responsible for setting up a limit–cycle instability (see §5.8 in Frank, King, & Raine 1992). Hachisu & Eriguchi (1983) have described a similar cycle of phase transitions in their/our Figures 1 and 2.

Notice also in Figure 1 that the one–ring sequence bifurcates smoothly at point A. The Maclaurin spheroid at A has a meridional eccentricity of $e$=0.985226 (Chandrasekhar 1967a) for which we find that $\omega_o^2$=8.72609, $j_o^2$=2.17412, and $E_o = -8.45757$ [in agreement with Bardeen's (1971) equivalent value of $j_o^2$=2.17410; see also Table 1 in §4.2 below]. Because the bifurcation is smooth, there is only one line tangent to the two sequences at A, which implies that the slope $d\omega/dj$ is the same for both sequences at A. This behavior probably occurs because equilibria along branch AB are not rings; they are "concave hamburgers," i.e. axisymmetric spheroids with a slight constriction at the ends of their symmetry axis (see Figure 2 in Eriguchi & Sugimoto 1981). We conclude that the "specific heat" $c_v \propto d\omega/dj$ (Paper I) is also continuous at the bifurcation point A. This smooth bifurcation prompted Hachisu & Eriguchi (1983) to identify the bifurcation point A with the appearance of a third–order phase transition. They then imagined that evolution proceeds along branch AB in Figure 1 and when point B is reached a dynamical transition takes the system down to point C.

However, evolution cannot proceed along branch AB in Figure 1 because, as is seen in Figure 2, equilibrium configurations along this branch have higher energies than the corresponding Maclaurin spheroids of the same angular momentum. Indeed, according to Figure 2 (and Figure 2 of Hachisu & Eriguchi 1983), all of the configurations along branches AB and BD and most of the configurations identified here along branch DC have higher energies than the corresponding Maclaurin spheroids. Hence, the third–order phase transition at point A is forbidden. In principle, a transition from the Maclaurin sequence to branch DC (along a vertical line in Figures 1 and 2) is allowed after branch DC crosses below the Maclaurin sequence in Figure 2, but branch BD continues to act as a free-energy barrier isolating the two minima of the free energy (represented by branches AX and DC) and impeding such a transition up to point X on the Maclaurin sequence.

### 2.3   The Free–Energy Function

In order to better understand the above description, we need to examine the behavior of the free–energy function in and out of equilibrium. We do this by imagining vertical "cuts" at constant values of $j_o$ in Figures 1 and 2. An example is shown in Figure 3. In the top two panels, we draw schematically plots similar to Figures 1 and 2 except that we assume for simplicity that branch DC crosses exactly at point A. (This assumption does not affect the results in any important way. Our description and conclusions remain unchanged even if branch DC crosses below the Maclaurin line prior to or past point A.) In the remaining six panels of Figure 3 we display schematically the free–energy function $E$ versus $\omega^2$ for different values of $j_o^2$. We begin with a vertical cut through point D and end with a cut past line XBC. Figure 3 is a powerful tool that immediately shows what takes place across the entire area of Figure 1. It also shows clearly what kinds of phase transitions appear between the two sequences and whether each of them is allowed or forbidden.

FIGURE 3. *In the top two panels, $\omega_o^2$ and $E_o$ in equilibrium are shown schematically versus $j_o^2$. In the top right diagram, we assume for simplicity that the branch DC crosses exactly at point A. In the next six panels, the free–energy function $E$ is shown schematically as a function of the (generally out-of-equilibrium) rotation frequency squared $\omega^2$ for different values of $j_o^2$. Each of the six panels is constructed by imagining a vertical "cut" at constant $j_o$ in the top two panels. Maxima, minima, and inflection points of $E$ are marked by filled circles. Letter symbols correspond to points discussed in the text. Numerical values correspond approximately to those in Figures 1 and 2.*

The roles that the different branches and points of the one–ring sequence play are easily understood from Figure 3: (a) Branch AB is composed of inflection points of the free–energy function. A system that finds itself anywhere along AB will return toward the Maclaurin sequence and will continue to execute stable oscillations about that free–energy minimum. (b) As we mentioned above, branch BD is composed of free–energy maxima, i.e. it is a branch of unstable equilibria. A system placed there may end up going to the Maclaurin or to the one–ring minimum depending on the relative depth of the two minima and on the specific characteristics of the applied perturbation. (c) Branch DC is the only branch of the one–ring sequence with stable equilibria. In fact, it represents the global minimum of the free–energy function past point A in Figure 3. (d) The Maclaurin line is also a branch of stable equilibria up to point X. (e) Point B lies at the intersection of the unstable branch BD and the "inflection" branch AB. It turns out that B represents an unstable equilibrium. (f) Point D lies at the intersection of the unstable branch BD and the stable branch DC. It turns out that D is an inflection point. (g) Point A in Figure 3 is a "critical point." The two free–energy minima have equal depths at A and thus the lower branch of the one–ring sequence becomes the new global minimum of the free–energy function past A.

Let us now return to Figure 1 keeping in mind the behavior of the free–energy function sketched in Figure 3. As we move from left to right at constant values of the dimensionless angular momentum, the Maclaurin line is the only available minimum prior to point D. At D, the one–ring sequence manifests itself by an inflection point at a lower value of $\omega_o$ relative to the Maclaurin line. This inflection point immediately splits into a minimum and a maximum representing the stable and the unstable branches of the one–ring sequence, respectively. Between D and B, unstable equilibria form a free–energy barrier separating the two stable minima. However, the global minimum still lies on the Maclaurin line between D and (roughly) A. Past the bifurcation point A, the "inflection" branch AB appears also and creates an inflection point between the Maclaurin line and the barrier. (Such inflection points are discussed in §3.2 and in the Appendix.) At a critical point (past but very near A in Figure 2; exactly at A in Figure 3), the two free–energy minima are equal. Past the critical point, the one–ring minimum is deeper than the Maclaurin minimum in which case this global minimum is also the "preferred" state (in the sense that a Maclaurin spheroid would like to be in that state of lower free–energy if it could make the transition). However, such a transition is still vigorously opposed by the free–energy barrier that separates the two states. The inflection point created by branch AB persists up to point B where it merges with the barrier. The barrier is still present at point B, however, and continues to block the transition to the lower energy state of the one–ring sequence.

## 2.4   The First–Order Phase Transition

We have stopped short of addressing the stability of the Maclaurin sequence past point X in order to make the following comments. What we have just described is a typical free–energy diagram of a first–order phase transition. In its simplest form (without an inflection point), such a diagram is characterized by two minima separated by a maximum in–between. We have described such a first–order phase transition in the past as a star–formation alter-native to dynamical collapse (Tohline 1985; Tohline, Bodenheimer, & Christodoulou 1987; Tohline & Christodoulou 1988). Unfortunately, we were not then aware of the results of Hachisu & Eriguchi (1983). Faithful to the energetics of first–order phase transitions in general, we continued that study by examining the magnitude of applied perturbations that is necessary in order for a system sitting at the shallow minimum to gain enough energy and to "jump up" to the top of the barrier and then down to the deep minimum (Christodoulou & Tohline 1990; Christodoulou, Sasselov, & Tohline 1993). We made a clear distinction between low–amplitude random perturbations applied continuously to a system over long periods of time and a strong, impulsive perturbation which would have the power to induce a "dynamical jump." In the former case, we showed that the barrier could be overcome after generally a long period of time and that such a first–order phase transition would only take place stochastically in the interstellar medium over long time scales. We called the latter case the "Jeans instability" because we imagined that a strong impulsive disturbance would displace the system to such a high energy state (higher than the top of the barrier itself) that it would no longer recognize its original equilibrium state or a free–energy barrier preventing its transition toward the deep, global free–energy minimum.

So, we did not appreciate fully the fact that the free–energy barrier itself can suddenly (i.e. discontinuously) disappear making the shallow minimum a saddle point and thereby causing a remarkable change in the appearance and the dynamics of the phase transition. But this is exactly what takes place in Figures 1 and 2 past the dotted line XBC. The last panel in Figure 3 clearly shows that, past point X, any point on the Maclaurin sequence is a saddle point in the free–energy function while a deep minimum exists on the stable toroidal branch of the one–ring sequence past point C. This transition looks like a typical *second–order* phase transition! This is quite unusual and implies that there is no degeneracy in the free–energy function between points X, B, and C. Instead, the free–energy function is not smooth across the transition line XBC (see also section A2 in the Appendix).

### 2.5    The λ–Transition to Torus

At any value of the angular momentum $j_o$ past the dotted line XBC in Figures 1 and 2 the transition between the Maclaurin and the one–ring sequence gives the impression of an apparently typical second–order phase transition in the free–energy function (see the last panel in Figure 3). In reality, it constitutes a remarkable new type of phase transition that does not obey the standard classification scheme of thermodynamical phase transitions (see e.g. Huang 1963). We describe its unusual features in this subsection and we discuss this transition in a more general context in sections A2 and A3 of the Appendix.

Recall the discussions in Paper I about second–order phase transitions. Such transitions appear beyond a point where a new sequence of equilibria bifurcates. They take place toward that sequence either dynamically or secularly depending on the (non)–conservation of the integrals of motion. A Maclaurin spheroid prior to the transition point occupies the only minimum of the free–energy function. Past the transition point, the spheroid finds itself on a saddle point while a descenting ridge leads downhill to a new state of lower free energy (see Figures 2 and 3 in Paper I). Because such transitions are related to the bifurcating sequences at the transition points, the derivative $dE/dj$ ("specific entropy") is continuous while the derivative $d\omega/dj$ ("specific heat") shows a finite discontinuity at each transition point. It is this finite discontinuity in the "specific heat" $c_v = I^2 d\omega/dj$ (where $I$ is the specific moment of inertia) that characterizes these transitions as second–order.

As we have discussed in §1.2 and as is seen in Figures 1 and 2, the transition that takes place past the dotted line XBC of Figure 1 does so in a "discontinuous" manner, in the sense that the one–ring sequence is not related to point X at all. In fact, the one–ring sequence is not related to any point on the Maclaurin line. Communication through the bifurcation point A is cut off because of the "barrier branch" BD seen in Figures 1 and 2. A Maclaurin spheroid prior to X may undergo a first–order phase transition and that only under very special conditions (a sufficiently strong, nonlinear perturbation that overcomes the barrier). Compared to that, communication between the two sequences is natural past point X thanks to the simultaneous disappearance of both branches AB and BD.

The transition past line XBC is "secular" because the following conditions are satisfied everywhere between the two equilibrium states (cf. the assumptions in §1.2 above): (a) Mass is conserved. (b) Evolution proceeds vertically in Figures 1 and 2 and at constant density and, thus, angular momentum is conserved. (c) For any specified value of the normalized angular momentum $j_o$ past line XBC there is no barrier separating the two states and the free energy of the toroidal equilibrium on the lower branch of the one–ring sequence is always smaller than the free energy of the corresponding unstable equilibrium on the Maclaurin sequence (Figure 2 and last panel in Figure 3). (d) Circulation is not conserved along the evolutionary path since viscosity redistributes the angular momentum efficiently and thus maintains uniform rotation.

The transition that takes place from any point on the Maclaurin sequence past X is vertical in Figures 1 and 2 and leads to a toroidal equilibrium at the corresponding point past C of the one–ring sequence. This results not only in a simple discontinuity in the specific heat but in $c_v \to \infty$, since $dj \to 0$ and $d\omega$ is finite at the transition point. The topology (i.e. the property of the volume of the spheroid to be simply–connected) is also "broken" by going from a spheroidal to a toroidal state. These characteristics are not observed during the occurrence of typical second–order phase transitions which show just a finite discontinuity in the specific heat and a breaking of the symmetry (Paper I). Infinite specific heat at the transition point has, however, been observed in the $\lambda$–transition of liquid $^4$He and in a few other transitions (e.g. the Curie–point transition in ferromagnetism and the order–disorder transition in binary alloys; see e.g. Huang 1963). The phase transition discussed above does not appear to be an order–disorder type of transition because no order is created and no symmetry breaks. We strongly suspect that this transition is the dynamical analog of the $\lambda$–transition of liquid $^4$He. For these reasons, we have adopted the name "$\lambda$–transition" instead of "apparent second–order" phase transition (see also section A3 in the Appendix).

### 3   FROM THE JACOBI TO THE DUMBBELL–BINARY SEQUENCE

#### 3.1   Previous Work

Bifurcation points along the incompressible Jacobi sequence corresponding to the third– and fourth–harmonics have been determined through the tensor virial method by Chandrasekhar (EFE and references therein). Eriguchi, Hachisu, & Sugimoto (1982) were able to calculate numerically the properties of both sequences, known as pear–shaped and dumbbell, that bifurcate from the Jacobi sequence at the corresponding points found by Chandrasekhar. Hachisu & Eriguchi (1984a, b) also calculated the binary sequence of two equal–mass detached objects as well as binary sequences with different mass ratios of the components. The equal–mass binary sequence merges with the dumbbell sequence and the two branches really form one continuous sequence that is usually called the dumbbell–binary sequence (see also more recent work by Hachisu 1986b). The point of contact of the two sequences represents an unstable equilibrium (a contact binary) because the free–energy function is a maximum (see Figure 5 below).

Detailed plots summarizing all the sequences that this group has calculated numerically can be found in Eriguchi & Hachisu (1983b, 1984). In these two papers, as well as in the two papers of Hachisu & Eriguchi (1983, 1984a), phase transitions between computed sequences have been discussed.

#### 3.2   The $\lambda$–Transition to Binary

The dumbbell–binary sequence bifurcates smoothly from the Jacobi sequence at a point where a fourth–harmonic mode of oscillation becomes neutral. The Jacobi ellipsoid at the bifurcation point has axes ratios of $b/a = 0.29720$ and $c/a = 0.25746$ (e.g. Chandrasekhar 1967b) corresponding to a meridional eccentricity of $e=0.96629$ and to an equatorial eccentricity of $\eta=0.95482$. For these values, we find that $\omega_o^2=5.32790$, $j_o^2=1.15082$, and $E_o = -6.23086$ (see also Table 1 in §4.2 below).

As Figures 4 and 5 indicate, the phase transitions between the Jacobi and the dumbbell–binary sequence are very similar to those described in §2. The schematic representation of the free–energy function (Figure 3) is also applicable here with one minor change. We see in Figure 5 that branch DC intersects the Jacobi line AX far from the bifurcation point A. In fact, point A has a lower value of $j_o$ than the lower turning point D. As a result, an inflection point appears in the vertical cuts past point A — i.e. prior to the inflection point at D — and persists throughout up to point B. At point B, this inflection point merges with the barrier just like in Figure 3. Past point B, the binary sequence identifies the only minimum in the free–energy function while the Jacobi ellipsoid sits on a saddle point (see also the last panel of Figure 3).

FIGURE 4. *The rotation frequency squared $\omega_o^2$ along the dumbbell–binary and the Jacobi sequence is plotted as a function of the dimensionless angular momentun squared $j_o^2$. The Jacobi sequence has been calculated analytically. The bifurcation point (filled circle) has also been calculated analytically using an equatorial axis ratio of $b/a = 0.29720$ (Chandrasekhar 1967b). All points (open circles) of the dumbbell–binary sequence other than the bifurcation point were obtained from Table II of Eriguchi, Hachisu, & Sugimoto (1982) and from Table 1 of Hachisu & Eriguchi (1984a). Letter symbols indicate points discussed in the text. The dotted line XBC denotes the onset of secular instability.*

All our discussions in §2 about a first–order phase transition prior to the dotted line XBC and about a $\lambda$–transition beyond that line are equally valid in the case plotted in Figures 4 and 5. The roles that different branches of the dumbbell–binary sequence play are also the same as before. Of course, no natural communication between mother and daughter sequence occurs through the bifurcation point A, i.e. the smooth bifurcation corresponds to a forbidden third–order phase transition here as well. The free–energy barrier is still present at point B and disappears discontinuously past line XBC implying again the the free energy is not a smooth function across the transition point. Finally, the "specific heat" past line XBC also becomes infinite during the $\lambda$–transition, symmetry to rotations by $\pm 180°$ is also preserved, and topology also "breaks" when an ellipsoid becomes a detached binary.

FIGURE 5. *The free energy $E_o$ is plotted versus $j_o^2$ for the sequences of Figure 4.*

We should mention that a vertical cut through D shows two inflection points both at higher free energies than the Jacobi minimum. The inflection points due to branch AB here and in §2 are probably due to the assumption of uniform rotation. These points do not split into a maximum and a minimum like points D in Figures 2 and 5 above. This may indicate that, by insisting on the assumption $\Omega =$ constant, we may have eliminated from consideration one or more sequences of two differentially rotating components (e.g. a core and a ring rotating at different frequencies; cf. Figure 3 in Eriguchi & Hachisu 1983a where such systems rotate with the same $\Omega$). Such sequences may be important in trying to understand the results from linear stability analyses (see Appendix). If considered, two–component sequences could transform the "inflection" branch AB into a typical "barrier" branch of unstable equilibria.

We have presented here a brief overview of the transitions between the Jacobi and the binary sequence because we believe that the $\lambda$–transition in Figures 4 and 5 plays a funda-mental role in connection with the classical fission problem of rotating fluids (see Lebovitz 1972, 1974, 1977, 1987; Tassoul 1978; Durisen & Tohline 1985). The $\lambda$–transition in Figure 4 is the last stage in a sequence of events that finally leads to fission of a quasistatically contracting system that evolves along the Jacobi sequence. Assuming that uniform rotation is maintained by viscosity, fission eventually occurs on a secular time scale just past point X in Figure 4. The Jacobi ellipsoid breaks up into two equal–mass fragments and the new-born binary system with the original symmetry preserved and the topology broken appears on the binary sequence of Figure 4 just past point C. Related to this scenario are also our beliefs that the slow evolution of a Jacobi ellipsoid does not proceed along the path of the pear–shaped sequence as Poincaré (1885) imagined and that no dynamical third–harmonic instability sets in at the Jacobi–pear bifurcation as was found by Cartan (1924). The physical and mathematical foundations that support this evolutionary scenario will be the principal subject of Paper III.

## 4   Discussion

### 4.1   Summary of Results

In this paper, we have examined in detail "fluid phase transitions" that appear between the Maclaurin and the one–ring sequence and between the Jacobi and the dumbbell–binary sequence. The one–ring and the dumbbell–binary (daughter) sequences bifurcate from the Maclaurin and the Jacobi (mother) sequences at points where linear stability analysis predicts the neutralization of an axisymmetric and a nonaxisymmetric fourth–harmonic mode of oscillation, respectively. The daughter sequences have very similar shapes (compare Figures 1 and 4) and dynamical/energetic manifestations (compare Figures 2 and 5). The characteristic "inverted S" shapes of these sequences that have two turning points in the angular momentum–rotation frequency plane are responsible for the appearance of three kinds of phase transitions: (a) Third–order phase transitions appear at the bifurcation points as the daughter sequences branch out smoothly into existence. These phase transitions are energetically forbidden because they lead to equilibria of higher energy. (b) First–order phase transitions appear in the general areas between the bifurcation points and the higher turning points. These transitions can take place only under the action of strong, nonlinear perturbations (see below). (c) "λ–transitions" appear past the higher turning points and are allowed to occur on secular time scales because they lead to states of lower free energy provided that uniform rotation is maintained along the evolutionary paths by the action of viscosity.

These three transitions have been identified through a careful examination of the free–energy function. The free–energy function for both pairs of sequences and for constant values of the total angular momentum has been illustrated schematically in Figure 3. The first–order phase transition is generally blocked by a barrier that separates the two free–energy minima corresponding to the mother and the daughter sequences. Hence, nonlinear perturbations are necessary to facilitate a transition in this case. Such strong perturbations may carry sufficient energy to displace a Maclaurin spheroid or a Jacobi ellipsoid oscillating about the shallow minimum over the top of the barrier and down to the deep minimum that describes a stable equilibrium of the lower branch of each daughter sequence. Related first–order phase transitions have been previously studied by Hachisu & Eriguchi (1983) and by our group (Tohline 1985; Christodoulou, Sasselov, & Tohline 1993 and references therein).

The daughter sequences turn back at the two turning points marked as B and D in Figures 1 and 4. As a result, there is no free–energy barrier past the higher turning point B and a phase transition is allowed to take place from saddle points representing now unstable equilibria on the mother sequences to the global free–energy minima on the stable lower branches of the daughter sequences. The transition is, however, secular because, owing to the assumption of uniform rotation, viscosity must be presumed to be present and sufficiently strong at all times. We call this secular instability the "$\lambda$–transition." The $\lambda$–transition occurs vertically, i.e. at constant angular momentum $j_o$, in the diagrams of Figures 1,2 and 4,5. As a result, the "specific heat" $c_v = I^2 d\omega/dj \to \infty$ at each transition point. Here, $I$ is the specific moment of inertia and $\omega$ is the rotation frequency. We believe that this transition is the dynamical analog of the notorious $\lambda$–transition of liquid $^4$He. The spatial symmetry of the Maclaurin spheroids (axisymmetry) and the Jacobi ellipsoids (symmetry to rotations by $\pm 180°$) does not break but the topology breaks in both cases during this transition. (This point is discussed in more detail in section A3 of the Appendix.) The breaking of topology is manifested as fission and is easily visualized in our systems: a Maclaurin spheroid becomes an axisymmetric torus and a Jacobi ellipsoid splits into a detached binary system.

FIGURE 6. *The ratio $T/|W|$ along the Maclaurin sequences of spheroids and toroids is plotted as a function of the dimensionless angular momentun squared $j_o^2$. The spheroidal sequence has been calculated analytically. All points (open circles) of the Maclaurin toroid sequence, including the bifurcation point (filled circle), were obtained from Eriguchi & Hachisu (1985). The point of dynamical axisymmetric instability (corresponding to e=0.99892) found by Bardeen (1971) coincides with the bifurcation point. Dynamical instability proceeds along the path denoted by a dotted line. Letter symbols indicate points discussed in the text.*

### 4.2   The Maclaurin Toroid Sequence

We describe here briefly the Maclaurin sequence of toroids which bifurcates from the Maclaurin sequence of spheroids in a different way than the one–ring sequence. The equilibrium properties of Maclaurin toroids were fully computed by Eriguchi & Hachisu (1985) and by Hachisu, Tohline, & Eriguchi (1987). The defining property of this sequence is that its members have exactly the same distribution of specific angular momentum as the Maclaurin spheroids of the same mass and total angular momentum. This means that circulation is also automatically conserved between the two sequences and that a dynamical phase transition should appear at a critical point. Therefore, Bardeen's point of dynamical instability must be related to this toroidal sequence as Eriguchi & Hachisu (1985) claimed originally. In fact, because of the particular way that the toroidal sequence bifurcates (see below), the point of dynamical instability found by Bardeen (1971) must also be the bifurcation point.

The Maclaurin sequences of spheroids and toroids are plotted in Figures 6 and 7 in the $(j_o^2, T/|W|)$ plane and in the $(j_o^2, E_o)$ plane, respectively, using data from Eriguchi & Hachisu (1985). The toroidal sequence bifurcates at point A where the spheroid has $e=0.99892$ corresponding to Bardeen's dynamical instability and to $j_o^2=4.59873$, $T/|W| = 0.45742$, and $E_o = -10.53929$. (Four points are seen slightly above the spheroidal sequence in Figure 6 and slightly below in Figure 7. This is a purely numerical artifact and helps to emphasize how smoothly this sequence bifurcates as well.)

FIGURE 7. *The free energy $E_o$ is plotted versus $j_o^2$ for the sequences of Figure 6.*

The phase transitions between the two sequences can be analyzed again by imagining vertical cuts at constant values of $j_o$ and by considering the behavior of the free–energy function $E$ along these cuts. Hachisu, Tohline, & Eriguchi (1987) nearly predicted the onset of the first–order phase transition when they argued that "nonlinear ring formation" should take place at point D in Figures 6 and 7. This is somewhat inaccurate since the first–order phase transition appears past D at point Y where the two minima of the free–energy function are equal (cf. the cut through point A in Figure 3 above). Point Y lies between D and the bifurcation point A and is characterized by the value $j_o^2$=3.149356. The Maclaurin spheroid at Y has $e$=0.995181, $T/|W|$=0.414195, and $E_o = -9.68913$. These values represent the exact point of "nonlinear ring formation" which can only be induced by strong nonlinear perturbations just as was discussed above for all the other first–order phase transitions.

Because the shape of the Maclaurin toroid sequence is different from the shape of the one–ring sequence, a vertical $\lambda$–transition is allowed precisely at the bifurcation point A in Figures 6 and 7 where the free–energy barrier disappears smoothly. Since the specific angular momentum is the same between the two sequences, all conservation laws are satisfied during this transition corresponding to dynamical instability. Thus, the bifurcation point A is also a point of dynamical instability (a $\lambda$–point) as was originally proposed by Eriguchi & Hachisu (1985). Owing to this coincidence, the "specific heat" is continuous at point A.

For convenience, we summarize in Table 1 all the relevant parameters of the bifurcation points, the critical points of the first–order phase transitions, and the $\lambda$–points of the $\lambda$–transitions discussed in this paper. The notation and the units of the different quantities are the same as in §1.2. The various types of $\lambda$–points are discussed in section A3 of the Appendix. The $\lambda$–transition to binary from the point of third–harmonic dynamical instability on the Maclaurin sequence (see §4.3 below and Paper IV) is also listed at the bottom of Table 1 for completeness. The entries noted as "EFE" were obtained using $e$=0.96696, the value for the third–harmonic dynamical instability on the Maclaurin sequence determined by linear stability analysis; the entries noted as "numerical" were obtained on the Maclaurin sequence using $j_o^2 = 1.540$, the value for the higher turning point of the dumbbell–binary sequence determined numerically by Eriguchi, Hachisu, & Sugimoto (1982); the differences between these two determinations are practically negligible.

The dynamical properties of the Maclaurin toroid sequence are helpful not only in understanding the results obtained from linear stability analyses but also in understanding the relation between the $\lambda$–transition of liquid [4]He and the Bose–Einstein condensation of an ideal Bose gas (see related discussions in the Appendix). This is because the fluid in these toroidal objects can be regarded as "noninteracting" since it has the same distribution of specific angular momentum (also circulation) as the original Maclaurin spheroids. On the other hand, the fluid in the tori of the one–ring sequence is "interacting" since the distribution of specific angular momentum is modified by viscosity that is assumed to be capable of preserving uniform rotation. The analogy to the ideal Bose gas of noninteracting particles and to liquid [4]He of interacting molecules, respectively, appears to be exact.

TABLE 1
BIFURCATION AND PHASE TRANSITION POINTS

### MACLAURIN SPHEROID TO ONE–RING TORUS

|                          | $e$       | $j_o^2$   | $\omega_o^2$ | $E_o$     | $T/|W|$  |
|--------------------------|-----------|-----------|--------------|-----------|----------|
| Bifurcation point        | 0.985226  | 2.174116  | 8.72609      | -8.45757  | 0.358906 |
| First–order critical point | 0.985599 | 2.195289 | 8.66330      | -8.49357  | 0.360444 |
| Type 2 $\lambda$–point   | 0.986834  | 2.270000  | 8.44183      | -8.61665  | 0.365731 |

### JACOBI ELLIPSOID TO BINARY

|                          | $\eta$   | $e$      | $j_o^2$   | $\omega_o^2$ | $E_o$    | $T/|W|$ |
|--------------------------|----------|----------|-----------|--------------|----------|---------|
| Bifurcation point        | 0.95482  | 0.96629  | 1.150820  | 5.32790      | -6.23086 | 0.18611 |
| First–order critical point | 0.96395 | 0.97203 | 1.318476  | 4.81362      | -6.91461 | 0.19367 |
| Type 2 $\lambda$–point   | 0.97201  | 0.97743  | 1.540000  | 4.25864      | -7.77253 | 0.20236 |

### MACLAURIN SPHEROID TO MACLAURIN TOROID

|                          | $e$       | $j_o^2$   | $\omega_o^2$ | $E_o$      | $T/|W|$  |
|--------------------------|-----------|-----------|--------------|------------|----------|
| Bifurcation & Type 1 $\lambda$–point* | 0.998917 | 4.598734 | 3.24702 | -10.53929 | 0.457424 |
| First–order critical point | 0.995181 | 3.149356 | 6.00990    | -9.68913   | 0.414195 |

### MACLAURIN SPHEROID TO BINARY

|                          | $e$       | $j_o^2$   | $\omega_o^2$ | $E_o$     | $T/|W|$  |
|--------------------------|-----------|-----------|--------------|-----------|----------|
| Type 3 $\lambda$–point (EFE) | 0.96696 | 1.537962 | 10.49141    | -7.10294  | 0.30308  |
| Type 3 $\lambda$–point (numerical) | 0.96705 | 1.540000 | 10.48661 | -7.10825 | 0.30329 |

* Also point of dynamical axisymmetric instability (Bardeen 1971).

### 4.3  Relation to Models of Evolving Protostellar Systems

Discussions of equilibrium sequences that bifurcate from the Maclaurin sequence of spheroids or the Jacobi sequence of ellipsoids have long been associated with theoretical scenarios that are designed to explain how stars, and in particular how binary star systems, form. (See Tassoul 1978, Durisen & Tohline 1985, and Bodenheimer 1992 for thorough reviews of this topic.) In Paper III, we shall discuss in considerable detail how the Jacobi-to-binary $\lambda$–transition is particularly expected to relate to the formation of binary stars. In brief, the transition provides a direct, energetically preferred evolutionary path between sequences of rapidly rotating, ellipsoidal figures and binary–star sequences. This path has not been heretofore examined. Previous numerical simulations of evolving protostars have not probed the parameter space that includes this pathway to binary star formation in part because the required initial (self–gravitating, ellipsoidal) configurations are difficult to construct for realistic equations of state; in part because fully three–dimensional simulations that properly incorporate the effects of kinematical viscosity are computationally demanding; and in part because previous theoretical scenarios have predicted that a third–harmonic instability will be encountered along the Jacobi sequence before the $\lambda$–point is reached. Evidence presented here and in Paper III that the $\lambda$–transition provides an energetically preferred route to "fission" should stimulate renewed interest in this topic.

The dynamical $\lambda$–transition of rapidly rotating, axisymmetric configurations from centrally condensed structures to ring–like (toroidal) configurations has been observed frequently in fluid dynamic simulations of protostellar collapse (Larson 1972; Black & Bodenheimer 1976; Nakazawa, Hayashi, & Takahara 1976; Tohline 1980; Boss & Haber 1982) as well as in simulations of stellar dynamic systems (Miller & Smith 1979). Such simulations clearly support the notion that, above some critical rotation, toroidal configurations provide lower–energy states to which protostellar clouds can evolve. Observationally, toroidal structures have been identified in gas clouds associated with star forming regions (Schloerb & Snell 1984; Zinnecker 1984; Schwartz et al. 1985). According to numerical simulations, such rings are usually unstable toward the development of nonaxisymmetric distortions which can lead to breakup into multiple (protostellar) fragments (cf., Tohline & Hachisu 1990, Christodoulou & Narayan 1992). Hence, understanding how rotating spheroids make a dynamical transition to toroidal configurations is also important to our basic understanding of how multiple star systems form. In this context, then, it is important to realize that that a secular $\lambda$–transition is also available to evolving spheroidal clouds if they are dominated by strong viscous processes. This $\lambda$–transition occurs at lower values of the total angular momentum providing the first direct avenue by which the transformation of a viscous Maclaurin spheroid to a viscous one–ring torus can take place.

Numerical simulations past the $\lambda$–point of the Jacobi sequence are, in some sense, analogous to the simulations performed past the second–harmonic point of dynamical instability on compressible Maclaurin sequences (e.g. Tohline, Durisen, & McCollough 1985; Williams & Tohline 1987, 1988). Such simulations attempted to examine a conjecture advanced by Ostriker (e.g. Ostriker & Bodenheimer 1973; see also Tassoul 1978) that dynamical fission may occur "cataclysmically" past that point. The results were negative because of efficient outward angular momentum transfer by spiral arms that eventually leads to ring formation around a remaining stable central object; this type of nonviscous evolution that is not constrained by uniform rotation thus leads finally to an equilibrium on some type of core–ring sequence of lower free energy compared to the $x = +1$ Riemann sequence (cf. Figure 3 in Eriguchi & Hachisu 1983a where similar sequences in uniform rotation are shown).

Our investigations reveal that there is "dynamical/cataclysmic" communication between the Maclaurin and the binary sequence but it occurs via a different type of phase transition not related to the second–harmonic dynamical instability (see Table 1). This different type of transition has been observed in the numerical simulations of Miyama, Hayashi, & Narita (1984) in which a rapidly rotating, spheroidal, isothermal cloud with $T/|W| = 0.3$ broke up into a binary (plus debris and a third, low–mass companion.) This cloud evolution is a demonstration of yet another astonishing $\lambda$–transition that originates on the Maclaurin sequence past the point of *third–harmonic dynamical instability* of meridional eccentricity $e \approx 0.967$ (see also Table 1 in §4.2 below). This $\lambda$–transition leads directly to the *binary* sequence while first symmetry and then topology break in sequence. The initial breaking of the symmetry (i.e. the existence of an unknown intermediate state of different symmetry and lower free energy which is not the global free–energy minimum) must be the reason that this type of dynamical $\lambda$–transition is detected by linear stability analysis. This evolutionary path will be analyzed in Paper IV in the context of third–harmonic perturbations applied to Maclaurin spheroids.

## Acknowledgments

We thank M. Ando, J. Brill, D. Brydges, B. Deaver, Jr., M. Elitzur, G. Hess, P. Mannheim, Q. Shafi, and J. Straley for stimulating discussions and I. Hachisu and N. Lebovitz for useful correspondence. We are especially grateful to C. McKee and to R. Narayan for in–depth discussions of the physical concepts presented in this paper. We are indebted to the graphics department of NRAO for assistance in producing Figure 3. IS is grateful to the Gauss Foundation for support and to K. Fricke, Director of Universitäts-Sternwarte Göttingen, for hospitality during a stay in which much of this work has been accomplished. IS thanks also the Center for Computational Studies of the University of Kentucky for continuing support. This work was supported in part by NASA grants NAGW–1510, NAGW–2447, NAGW–2376, and NAGW–3839, by NSF grant AST–9008166, and by grants from the San Diego Supercomputer Center and the National Center for Supercomputing Applications.

APPENDIX
## A Discussion of Related Topics

### A1. *The Results from Linear Stability Analyses*

A comparison between our results and the corresponding results from linear stability analyses (e.g. Chandrasekhar 1967a,b, 1971; Bardeen 1971) leads to the following conclusions. (a) Linear analyses missed the secular instability represented by the $\lambda$–transitions at the higher turning points of the one–ring and the dumbbell–binary sequences. (b) Linear analyses did not detect the first–order phase transitions. (c) The dynamical axisymmetric instability found by Bardeen (1971) at $e$=0.99892 does occur at the bifurcation point of the Maclaurin toroid sequence. (d) Linear analysis predicts a neutral mode (and thus secular instability) only at the bifurcation points of the daughter sequences. We have seen above that third–order phase transitions at the same points leading to uniformly rotating equilibria are energetically forbidden.

The first three points are the only ones that are clearly understood. First–order phase transitions and "secular" $\lambda$–transitions are fully nonlinear effects and cannot be detected by any linear stability technique. The "dynamical" $\lambda$–transition was detected by linear analysis because it occurs at a bifurcation point and all conservation laws are automatically satisfied (§4.2). Even in this case, however, the transition is vertical in Figures 6 and 7 and does not proceed along the bifurcating sequence which initially unfolds toward higher energies relative to the mother sequence.

All $\lambda$–transitions can be interpreted as nonlinear catastrophes. The $\lambda$–points are saddle points of the free–energy function $E$ implying that the gradient with respect to the coordinates $x_i$ is zero, i.e. $\nabla \mathrm{E} = 0$ and, in addition, that the determinant of the Hessian matrix $E_{ij} \equiv \partial^2 E / \partial x_i \partial x_j$ is also zero, i.e. $det(E_{ij}) = 0$. Linear analysis then fails to discover some of the $\lambda$–transitions because this matrix is not invertible and standard Taylor expansions are invalid in the vicinity of $\lambda$–points. Thom's (1975) *elementary catastrophe theory* provides a means of obtaining the "canonical form," i.e. a meaningful expansion around $\lambda$–points, if the free energy is a smooth function and generally depends on no more than five integrals of motion (see e.g. Gilmore 1981). The phase transitions described by Landau & Ginzburg (Landau & Lifshitz 1986) not only fit naturally within the framework of catastrophe theory but are also better understood physically as well. On the other hand, the "secular" $\lambda$–transitions are more complicated as is briefly discussed in the context of catastrophe theory in section A2 below.

The differences in point (d) above are not easily explained. Also related to these differences are the following results obtained by Chandrasekhar (1967b) and Bardeen (1971) who performed linear analyses in the post–Newtonian approximation. Singularities in the post–Newtonian solutions appeared at the bifurcation points of the one–ring and the dumbbell–binary sequences but not at the Maclaurin–Jacobi bifurcation, at the Jacobi–pear bifurcation, or at the point $e$=0.99892 of Bardeen's dynamical instability. (Bardeen wrote by mistake that a singularity occurs at the Jacobi–pear bifurcation.) Such singularities in the post–Newtonian solutions were interpreted by Chandrasekhar and Bardeen as excitations of the corresponding Newtonian secular instabilities by the post–Newtonian displacements that occur when these displacements have the proper symmetry (axisymmetric and cubic) and do not violate equilibrium conditions.

The post–Newtonian singularities and the Newtonian secular instabilities at the bifurcation points are not clearly understood. Singular solutions were found only in the two cases where "inflection" branches appear (branches AB in Figures 1 and 4). As we discussed in §3.2, these inflection points may imply that additional sequences of differentially rotating two–component equilibria have not been considered in each case (cf. Bardeen 1971). It is thus conceivable that one such sequence represents a free–energy minimum leaving the original states on saddle points past the bifurcation points. An alternative is that the "inflection" branches themselves may have lower free energies than the mother sequences in the post–Newtonian approximation. In either case, Chandrasekhar's and Bardeen's interpretation of the post–Newtonian results appears to be correct.

Consider now the Newtonian neutral modes found by linear analyses in conjunction with additional bifurcating sequences of differentially rotating equilibria. In such a case, the first–order perturbation theory used should suffer from degeneracy at the bifurcation points. It is known that the free energy changes to second order as the rotation frequency deviates from uniform rotation (Bardeen 1971 and references therein). A neutral mode may then mean that both the one–ring/dumbbell sequence and a differentially rotating two–component sequence branch off at the same bifurcation point. Thus, second–order perturbation theory is needed to detect the splitting to a local maximum (the "inflection" branch) and a local minimum (the two–component sequence) away from each bifurcation point.

Consider next the "inflection" branches AB in Figures 1, 4 and 2, 5 which represent local maxima in the picture just described. In a third–order phase transition, both the first $dE/dj$ and the second $d\omega/dj$ derivatives of $E$ are continuous. The continuity of $d\omega/dj$ is clearly seen in Figures 1 and 4. In addition, the slopes $dE/dj$ should also be continuous at the bifurcation points A in Figures 2 and 5. Such continuity is probably seen in Figure 5. In Figure 2, no points close to A were listed in the table of models that we have used. So, the behavior of the "inflection" branches does not contradict the above picture of degeneracy at each bifurcation point. Note, however, that yet another axisymmetric sequence would have to be positioned between the "inflection" branch and the "barrier" branch to account for a local minimum in–between these two free–energy maxima. (An alternative would be the turning of the two–component sequence itself exactly at the higher turning point B creating another minimum between branches AB and BD in Figures 1 and 4.)

If the above picture is correct, then we may understand the Newtonian secular instabilities at the bifurcation points since circulation will not be conserved between each mother sequence and its daughter two–component sequence. Hence, the transition from the former to the latter sequence would only be allowed to proceed secularly in the presence of viscosity very much like the Maclaurin–Jacobi secular transition discussed in Paper I and like the $\lambda$–transitions of §2 and §3. In the same picture, post–Newtonian effects would alter the character of the instability because, in addition, they violate the conservation law of angular momentum.

A2. *Structural Stability and Catastrophic Morphogenesis*

We have seen above that nonlinear processes, like the first–order phase transitions and the "secular" $\lambda$–transitions of §2 and §3, are not detected by linear stability analyses. The desire to study such nonlinear processes has led to the development of the concept of *structural stability* in the context of *catastrophe theory* which provides nonlinear methods generally more powerful and more insightful than older methods such as the Landau second–order stability analysis and bifurcation theory (see §49 and §52 in Drazin & Reid 1981).

A system is called structurally unstable if infinitesimal perturbations cause dramatic changes to its behavior and its stability. Drazin & Reid (1981, §52.3) offer the stability properties of "perfect" and viscous fluids as a typical example of structural instability. In many cases, introduction of an infinitesimal amount of viscosity in the Euler equations of motion of a perfect fluid causes a dramatic change to the stability of the fluid. Furthermore, taking the equations in the limit of no viscosity does not always recover the results obtained for a perfect fluid. Drazin & Reid argue that this inconsistency represents a limitation of the theory of inviscid fluids. We believe that this example does not really demonstrate any inconsistency. As we have discussed in detail in Paper I and in §1 above, viscous and inviscid fluids are fundamentally different irrespective of the magnitude of viscosity present in the former. This fundamental difference is related to the conservation law of circulation. Circulation is not conserved in viscous fluids. As a result, the evolution and stability properties of a viscous fluid are constrained by one less conservation law and there probably lies the explanation for the observed different results. Consider, for example, second–harmonic perturbations applied to Maclaurin spheroids. We have seen in Paper I that constraining the free–energy function by the conservation law of circulation renders the point of secular instability irrelevant and dynamical instability appears at a distant point, the bifurcation point of the $x = +1$ Riemann sequence. The same result can be obtained by letting the coefficient of kinematic viscosity $\nu \to 0$ in the solution for the oscillation frequencies of the viscous Maclaurin spheroid (see §37c in EFE) but only after accounting for a singularity in the solution at the point of dynamical instability (cf. section A1). We have also seen in Paper I a related dramatic difference to the character of the instability of stellar systems at the Maclaurin–Jacobi bifurcation. In stellar systems, this instability is dynamical because circulation is destroyed on dynamical time scales. The above examples are typical manifestations of structural instability in viscous/inviscid systems. They indicate a direct connection between structural instability and the presence or the absence of a conservation law. We have conjectured in section A1 that an analogous case of structural instability occurs when Newtonian secular instabilities change their character in the post–Newtonian approximation.

The $\lambda$–transitions from the Maclaurin sequence also provide an example of structural instability: imposing circulation conservation renders the $\lambda$–transition toward the one–ring sequence irrelevant and dynamical axisymmetric instability appears at the bifurcation point of the Maclaurin toroid sequence. Furthermore, this dynamical instability and the transition in Tohline (1985) are both typical cases of "imperfect supercritical instability" that is discussed by Drazin & Reid (1981; §52.3) in the context of Thom's (1975) catastrophe theory. [See also the clear presentation of the application of this theory to phase transitions by Gilmore (1981).] Imperfect supercritical instability is usually demonstrated on a folded (more precisely "cusped") surface in three–dimensional space. An evolving system on the higher level of the surface that reaches the edge of the fold drops down to the lower level of the same surface. The abrupt fall over the edge is called a catastrophe.

The $\lambda$–transition to the Maclaurin toroid sequence that occurs at point A in Figures 6 and 7 and the $\lambda$–transition in Figures 2 and 3 of Tohline (1985) represent such a catastrophe. These transitions are characterized by a smooth disappearance of the free–energy barrier and by a degeneracy of the free–energy function at the transition point. The degeneracy is lifted past the transition point and a new minimum of the free energy appears. Because the free–energy function exhibits smooth variations such transitions can be described by one of Thom's (1975) elementary catastrophes, the so–called "cusp catastrophe." Depending on the particular chosen evolutionary path, the cusp catastrophe exhibits either only a $\lambda$–transition or a first–order phase transition followed by a $\lambda$–transition or only a typical second–order phase transition of the kind discussed in Paper I. The first case that is relevant to the $\lambda$–transition between the Maclaurin and the so–called two–ring sequence (Eriguchi & Hachisu 1982, 1983a) will be described in Paper IV; the second case is described by Gilmore (1981, §10.2); and the third case has been analyzed by Poston & Stewart (1978, §14.16) using the results of Bertin & Radicati (1976).

In contrast, the $\lambda$–transitions of §2 and §3 and, by analogy, the $\lambda$–transition of liquid $^4$He are not elementary catastrophes because the free–energy barrier disappears discontinuously at the transition point (see e.g. Figure 3). Thus, the transition point is not a degenerate critical point; instead, it represents a discontinuity in the free–energy function. Such a discontinuity cannot be described by an elementary catastrophe since catastrophe theory describes the dynamics of families of only smooth functions. [We note, however, that Keller, Dangelmayr, & Eikemeier (1979) have argued that the $\lambda$–transition of liquid $^4$He may be a part of the so–called "butterfly catastrophe."] For this reason, we believe that the results of §2 and §3 and Figure 3 indicate that an extension of catastrophe theory to families of discontinuous functions is needed to describe the discontinuous $\lambda$–transitions of §2 and §3 and the macroscopic manifestations of the superfluid $^4$He transition. This subject is actually quite technical and involved and we are not going to pursue it further in this series of papers.

The catastrophe that occurs during any type of $\lambda$–transition always leads to breaking of the topology of the original system. Topology breaking is the only common characteristic of all types of $\lambda$–transitions since the "specific heat" may or may not diverge at the $\lambda$–point (see §2 and §4.2) and symmetry may or may not break (see section A3 below). In all cases studied above, the form (i.e. the structure) of the newborn system is quite different visually as well as topologically. We have thus a clear and complete visual/physical picture of the phenomenon of *morphogenesis* (i.e. the creation of new forms or figures like tori and binaries) after the catastrophes that occur during all types of $\lambda$–transitions in our systems: Geometrically, catastrophic morphogenesis is always the result of the breaking of topology; physically, it is achieved on various time scales via secular or dynamical $\lambda$–transitions depending on which conservation laws are valid during evolution. (We should emphasize again that neither the behavior of the specific heat nor the resulting symmetry is the primary characteristic of catastrophic morphogenesis.)

### A3. $\lambda$–Transitions and Topology Breaking

We have analyzed above two out of four distinct types of $\lambda$–transitions with the following properties. (a) The "specific heat" $c_v = I^2 d\omega/dj \to \infty$ in the $\lambda$–transitions of §2 and §3 but remains continuous in the $\lambda$–transition of §4.2 because this transition occurs precisely at the bifurcation point of the Maclaurin toroid sequence. (b) Topology breaks in both cases. (c) Symmetry does not break in either of these two cases. (The *third type* of $\lambda$–transition discussed briefly in §4.3 results in breakings of first the symmetry and then the topology and will be analyzed in Paper IV along with a *fourth type* which preserves the symmetry but does not allow for a preceding first–order transition.)

The above two $\lambda$–transitions with no symmetry breaking are the dynamical analogs of the $\lambda$–transition of liquid $^4$He and of the Bose–Einstein (in brief, BE) condensation of an ideal Bose gas (see e.g. London 1954; Huang 1963; Landau & Lifshitz 1986).

Consider first the BE condensation. It occurs below a critical temperature in momentum space when a finite fraction of the noninteracting particles occupies the zero–momentum level. All particles have zero momentum at the temperature of the absolute zero while, in contrast, they are all "thinly spread" over all available levels above the transition temperature, with no level having a finite occupation number. There is clearly some confusion in the literature about the thermodynamical manifestations of this transition. Huang (1963) classifies it as a typical first–order phase transition because latent heat is released (see the Appendix in Paper I). Atkins (1959) calls it a third–order phase transition because the specific heat is continuous at the transition point.

Consider next the $\lambda$–transition of liquid $^4$He. Huang (1963) explains that this transition does not fit the standard classification scheme of thermodynamical phase transitions because the specific heat $c_v$ diverges at the $\lambda$–point not allowing for a determination of the latent heat $\mathcal{L}$ from the equation $\mathcal{L} = c_v d\theta$ (Paper I), where $\theta$ denotes temperature. (Technically, Huang claims that there is no latent heat and none has been measured yet. However, at the $\lambda$–point, $d\theta \to 0$ as $c_v \to \infty$ but their product $\mathcal{L}$ may tend to a small but finite value.) Atkins (1959) explains that the $\lambda$–transition is manifested as a second–order phase transition according to second–order perturbation theory based on an approximate wavefunction devised by Feynman & Cohen (1956). This is similar to the second–order phase transition found by first–order perturbation theory in an imperfect Bose gas (e.g. Huang 1963) when weak repulsive interactions between particles are approximately included. It is also found that this spatial repulsion leads to "momentum–space attraction," an effect that is in the right direction considering that even stronger intermolecular interactions cause a momentum–space catastrophe, i.e. a discontinuous $\lambda$–transition (see below).

We may be able to understand macroscopically both the above thermodynamical transitions by using what we have learned from the "$\lambda$–transitions" of our dynamical systems. Consider first the $\lambda$–transition to the Maclaurin toroid sequence. This transition occurs exactly at the bifurcation point A in Figure 6. It is a vertical transition at constant angular momentum and thus "latent heat" $\mathcal{L} = d\omega^{-1}$ (Paper I) is released. But the "specific heat" $c_{\mathrm{v}} = I^2 d\omega/dj$ is continuous at the transition point because the daughter sequence bifurcates smoothly in Figure 6. (The bifurcation is backwards and the $\lambda$–transition is not blocked by a free–energy barrier.) This type of $\lambda$–transition with continuous $c_{\mathrm{v}}$ and positive $\mathcal{L}$ is exactly analogous to the BE condensation of an ideal Bose gas. Based on this analogy, we conclude that the BE condensation is not a first– or a third–order phase transition; it is the *first type* of a $\lambda$–transition analogous to that highlighted by the dotted lines in Figures 6 and 7 above. Only topology breaks in the hydrodynamical transition while axisymmetry is preserved. Similarly, we imagine that only momentum–space (i.e. phase–space) topology breaks in the BE condensation since the ideal Bose gas is so simple with no available symmetries to be broken.

Consider now the $\lambda$–transition of liquid $^4$He in relation to both the BE condensation and the "$\lambda$–transition" in Figure 6. If the superfluid transition of $^4$He is some kind of BE condensation modified by intermolecular interactions (as is widely believed), then at least two observations (e.g. Huang 1963) should be explained with the help of Figure 6: (a) the specific heat diverges at the $\lambda$–point of liquid $^4$He; and (b) the $\lambda$–point of liquid $^4$He occurs at $\theta = 2.18$ K, i.e. at a lower temperature than $\theta = 3.14$ K of the BE condensation point of an ideal Bose gas with the same particle mass and volume density as liquid $^4$He.

Both of these observations can be explained by assuming that the analogous dynamical effect of strong intermolecular interactions is to invert and to bend again the daughter sequence in Figure 6, making it look like the "inverted S" sequences in Figures 1 and 4. This is the only way to prevent the $\lambda$–transition prior to and at the bifurcation point. However, the resulting shape now allows for a $\lambda$–transition past the higher turning point. This *second type* of $\lambda$–transition then naturally has $c_{\mathrm{v}} \to \infty$ because it no longer occurs at the bifurcation point and naturally occurs at a somewhat lower temperature (i.e. at a higher $j_o$ value in Figure 6; see the Appendix in Paper I), just as is observed for liquid $^4$He.

The assumption of sequence inversion due to the presence of molecular interactions is reasonable. Such an inversion may occur close to the bifurcation point since interacting molecules obey different equilibrium conditions than their noninteracting counterparts of an ideal Bose gas. A combination of repulsive and attractive interaction terms may create an unstable "inflection" branch that extends to lower temperatures (as in Figures 1 and 4) but eventually will have to give in and to allow the equilibrium sequence to turn back toward higher temperatures as in the case of noninteracting molecules. It is such strong interactions that give rise to the "roton" excitations in liquid $^4$He in addition to the "phonons" that exist alone at very low temperatures (e.g. Huang 1963). The assumption of another bending of the sequence back toward lower temperatures is also reasonable. The equilibrium sequence must eventually bend back toward low temperatures in order to terminate at the absolute zero (which is at an even lower temperature than the new $\lambda$–point). This corresponds to the disappearance of the "roton" branch from the dispersion relation of liquid $^4$He at very low temperatures where intermolecular interactions diminish in importance.

Gilmore (1981, §18.5) gives a similar description of sequence inversion and bending for bifurcating sequences in the context of catastrophe theory. The only difference between the two descriptions is that Gilmore's illustration is just a general mathematical model while the above description applies specifically to the macroscopic physics of liquid $^4$He and the BE condensation.

The above classification of hydrodynamical $\lambda$–transitions into four *types* has an immediate impact on the standard classification scheme of thermodynamical phase transitions. The $\lambda$–transitions should be introduced as a separate class with the understanding that this class includes four distinct types (see also Paper IV) linked together by the breaking of topology. It is also clear from the above discussion that the order–disorder transition in binary alloys like $\beta$–brass (§2.5; Huang 1963) appears to be analogous to the third type of symmetry–breaking $\lambda$–transition mentioned in this subsection and in §4.3. Finally, the Curie–point transition in ferromagnetism (Huang 1963) is part of the cusp catastrophe (Poston & Stewart 1978, §14.2) and appears to be analogous to the fourth type of $\lambda$–transition discussed in Paper IV in relation to the two–ring sequence (Eriguchi & Hachisu 1982, 1983a).

REFERENCES

Atkins, K. R. 1959, Liquid Helium (Cambridge: Cambridge Univ. Press)
Bardeen, J. M. 1971, ApJ, 167, 425
Bertin, G., & Radicati, L. A. 1976, ApJ, 206, 815
Black, D. C., & Bodenheimer, P. 1976, ApJ, 206, 138
Bodenheimer, P. 1992, in Star Formation in Stellar Systems, ed. G. Tenorio–Tagle *et al.* (Cambridge: Cambridge Univ. Press), 1
Boss, A. P., & Haber, J. G. 1982, ApJ, 255, 240
Cartan, H. 1924, Proc. Int. Math. Congress, Toronto 1928, 2, 2
Chandrasekhar, S. 1967a, ApJ, 147, 334
Chandrasekhar, S. 1967b, ApJ, 148, 621
Chandrasekhar, S. 1969, Ellipsoidal Figures of Equilibrium (New Haven: Yale Univ. Press), (EFE)
Chandrasekhar, S. 1971, ApJ, 167, 455
Christodoulou, D. M., Kazanas, D., Shlosman, I., & Tohline, J. E. 1994, ApJ, submitted (Paper I)
Christodoulou, D. M., Kazanas, D., Shlosman, I., & Tohline, J. E. 1994, ApJ, in preparation (Paper III)
Christodoulou, D. M., Kazanas, D., Shlosman, I., & Tohline, J. E. 1994, ApJ, in preparation (Paper IV)
Christodoulou, D. M., & Narayan, R. 1992, ApJ, 388, 451
Christodoulou, D. M., Sasselov, D. D., & Tohline, J. E. 1993, ASP Conf. Ser., 45, 365
Christodoulou, D. M., & Tohline, J. E. 1990, ApJ, 363, 197
Constantinescu, D. H., Michel, L., & Radicati, L. A. 1979, Le Journal de Physique, 40, 147
Drazin, P. G., & Reid, W. H. 1981, Hydrodynamic Stability (Cambridge: Cambridge Univ. Press)
Durisen, R. H., & Tohline, J. E. 1985, in Protostars and Planets II, ed. D. C. Black & M. S. Matthews (Tucson: Univ. of Arizona Press), 534
Eriguchi, Y., & Hachisu, I. 1982, Prog. Theor. Phys., 67, 844
Eriguchi, Y., & Hachisu, I. 1983a, Prog. Theor. Phys., 69, 1131
Eriguchi, Y., & Hachisu, I. 1983b, Prog. Theor. Phys., 70, 1534
Eriguchi, Y., & Hachisu, I. 1984, PASJ, 36, 491
Eriguchi, Y., & Hachisu, I. 1985, AA, 148, 289
Eriguchi, Y., Hachisu, I., & Sugimoto, D. 1982, Prog. Theor. Phys., 67, 1068

Eriguchi, Y., & Sugimoto, D. 1981, Prog. Theor. Phys., 65, 1870

Feynman, R. P., & Cohen, M. 1956, Phys. Rev., 102, 1189

Frank, J., King, A. R., & Raine, D. J. 1992, Accretion Power in Astrophysics (Cambridge: Cambridge Univ. Press)

Gilmore, R. 1981, Catastrophe Theory for Scientists and Engineers (New York: Dover)

Hachisu, I. 1986a, ApJS, 61, 479

Hachisu, I. 1986b, ApJS, 62, 461

Hachisu, I., & Eriguchi, Y. 1982, Prog. Theor. Phys., 68, 206

Hachisu, I., & Eriguchi, Y. 1983, MNRAS, 204, 583

Hachisu, I., & Eriguchi, Y. 1984a, PASJ, 36, 239

Hachisu, I., & Eriguchi, Y. 1984b, PASJ, 36, 259

Hachisu, I., & Eriguchi, Y. 1984c, PASJ, 36, 497

Hachisu, I., Eriguchi, Y., & Sugimoto, D. 1982, Prog. Theor. Phys., 68, 191

Hachisu, I., Tohline, J. E., & Eriguchi, Y. 1987, ApJ, 323, 592

Huang, K. 1963, Statistical Mechanics (New York: Wiley)

Keller, K., Dangelmayr, G., & Eikemeier, H. 1979, in Structural Stability in Physics, ed. W. Güttinger & H. Eikemeier (Berlin: Springer–Verlag), 186

Lamb, H. 1932, Hydrodynamics (New York: Dover)

Landau, L. D., & Lifshitz, E. M. 1986, Statistical Physics, Part 1 (New York: Pergamon Press)

Larson, R. B. 1972, MNRAS, 156, 437

Lebovitz, N. R. 1972, ApJ, 175, 171

Lebovitz, N. R. 1974, ApJ, 190, 121

Lebovitz, N. R. 1977, in Applications of Bifurcation Theory (New York: Academic Press), 259

Lebovitz, N. R. 1987, Geophys. Astrophys. Fl. Dyn., 38, 15

London, F. 1954, Superfluids, Vol. II (New York: Wiley)

Lyttleton, R. A. 1953, The Stability of Rotating Liquid Masses (Cambridge: Cambridge Univ. Press)

Miller, R. H., & Smith, B. F. 1979, ApJ, 227, 407

Miyama, S. M., Hayashi, C., & Narita, S. 1984, ApJ, 279, 621

Nakazawa, K., Hayashi, C., & Takahara, M. 1976, Prog. Theor. Phys., 56, 515

Ostriker, J. P., & Bodenheimer, P. 1973, ApJ, 180, 171

Poincaré, H. 1885, Acta Math., 7, 259

Poston, T., & Stewart, I. N. 1978, Catastrophe Theory and its Applications (London: Pitman)

Schloerb, F. P., & Snell, R. L. 1984, ApJ, 283, 129

Schwartz, P. R., Thronson, H. A., Jr., Odenwald, S. F., Glaccum, W., Loewenstein, R. F., & Wolf, G. 1985, ApJ, 292, 231

Tassoul, J.-L. 1978, Theory of Rotating Stars (Princeton: Princeton Univ. Press)

Thom, R. 1975, Structural Stability and Morphogenesis (Reading: Benjamin)

Tohline, J. E. 1980, ApJ, 236, 160

Tohline, J. E. 1985, ApJ, 292, 181

Tohline, J. E., Bodenheimer, P. H., & Christodoulou, D. M. 1987, ApJ, 322, 787

Tohline, J. E., & Christodoulou, D. M. 1988, ApJ, 325, 699

Tohline, J. E., Durisen, R. H., & McCollough, M. 1985, ApJ, 298, 220

Tohline, J. E., & Hachisu, I. 1990, ApJ, 361, 394

Wilks, J., & Betts, D. S. 1987, An Introduction to Liquid Helium (Oxford: Clarendon Press)

Williams, H. A., & Tohline, J. E. 1987, ApJ, 315, 594

Williams, H. A., & Tohline, J. E. 1988, ApJ, 334, 449

Zinnecker, H. 1984, Ap. Space Sci., 99, 41